\documentclass[%
reprint,
superscriptaddress,
showpacs,preprintnumbers,
amsmath,amssymb,
prb
]{revtex4-1}

\usepackage{graphicx}
\usepackage{dcolumn}
\usepackage{bm}

\usepackage{color} 

\begin{document}

\preprint{APS/123-QED}

\title{Ultrafast asymmetric Rosen-Zener-like coherent phonon responses observed in silicon
}

\author{Yohei Watanabe}
\affiliation{Doctoral Program in Materials Science, Graduate School of Pure and Applied Sciences, University of Tsukuba, Tsukuba, Ibaraki 305-8573, Japan}
\author{Ken-ichi Hino}
\email{hino@ims.tsukuba.ac.jp}
\affiliation{Division of Materials Science, Faculty of Pure and Applied Sciences, University of Tsukuba, Tsukuba 305-8573, Japan}
\affiliation{Center for Computational Sciences, University of Tsukuba, Tsukuba 305-8577, Japan}
\author{Nobuya Maeshima}
\affiliation{Center for Computational Sciences, University of Tsukuba, Tsukuba 305-8577, Japan}
\affiliation{Division of Materials Science, Faculty of Pure and Applied Sciences, University of Tsukuba, Tsukuba 305-8573, Japan}
\author{Hrvoje Petek}
\affiliation{Department of Physics and Astronomy and Pittsburgh Quantum Institute, University of Pittsburgh, Pittsburgh, Pennsylvania 15260, USA}
\author{Muneaki Hase}
\affiliation{Division of Applied Physics, Faculty of Pure and Applied Sciences, University of Tsukuba, Tsukuba 305-8573, Japan}

\date{\today}

\begin{abstract}
We investigate the spectral profiles of time signals attributed to coherent phonon generation in an undoped Si crystal.
Here, the retarded longitudinal-optical (LO) phonon Green function relevant to the temporal variance of induced charge density of ionic cores is calculated by employing the polaronic quasiparticle model developed by the authors [Y. Watanabe {\it et al}., Phys. Rev. B {\bf 95}, 014301 (2017); {\it ibid}., {\bf 96}, 125204 (2017)].
The spectral asymmetry is revealed in the frequency domain of the signals under the condition that an LO phonon mode stays almost energetically resonant with a plasmon mode in the early time region; this lasts for approximately 100 fs immediately after the irradiation of an ultrashort pump-laser pulse.
It is understood that based on the adiabatic picture in time, this asymmetry is caused by the Rosen-Zener coupling between both  modes.
The associated experimental results are obtained by measuring time-dependent electro-optic reflectivity signals, and it is proved that these are in harmony with the calculated ones.
The spectra become more symmetric, as the photoexcited carrier density further changes from that meeting the above condition to higher and lower sides of carrier densities.
Moreover, the effect of optical nutation of carrier density on the CP signals is addressed, and the present results are compared with the asymmetry caused by transient Fano resonance, and the spectral profiles observed in a GaAs crystal in the text.

\end{abstract}

\maketitle


\section{INTRODUCTION}
\label{sec1}
Using ultrafast laser pulse irradiation to semiconductors, it is possible to impulsively generate bare particles, e.g., phonons, and observe their subsequent dressing by the many-body interactions, that is, formation of a quasiparticle.~\cite{Huber1,hase1} 
The use of quasiparticles in semiconductors may revolutionize modern semiconductor-based optical and electronic devise technologies, such as quantum computing.~\cite{Leitenstorfer1,Jurcevic1}
Coherent phonon (CP) generation~\cite{kuznetsov1} induced by an ultrashort pulse laser has been investigated in various materials,~\cite{cho1, pfeifer1, hase1, riffe1, hase2, Basak1, yoshino1,ishioka1, hase3, misochko1, ishioka2, li1,chwalek1, albrecht1, misochko2} and efforts of exploration in this research area have been devoted to the understanding of the underlying microscopic mechanism governing the CP generation~\cite{scholz1, nakamura1, watanabe1, watanabe2} and the concomitant quantum mechanical effects just after the pulse-irradiation.~\cite{hase1, yoshino1, hase3} 
The experimental results of the CP generation have been examined on the basis of the two well-known classical models subject to a damped forced-oscillation; the impulsive stimulated Raman scattering model~\cite{yan1} and the displacive excitation of CP model.~\cite{zeiger1}
These models succeed in part in demonstrating the qualitative features of the CP oscillation.~\cite{riffe1, garrett1, stevens1, bragas1}
However, such an approach is confronted with difficulties in describing the early stage of the CP generation dynamics that is governed by an interaction of strongly photoexcited carriers with longitudinal optical (LO) phonons. 
Hereafter, this stage is termed as the early time region (ETR), which lasts up to approximately 100 fs after the irradiation of pump-pulse.
Actually, it is commonly observed in the ETR in experiments that oscillatory patterns of transient electro-optic reflectivity signals for the CP are largely deviated from 
signals observed in the temporal region following the ETR; this is termed hereafter as the classical region.

Such anomalous signals sharply distinguished from a damped harmonics
are considered to result from certain quantum-mechanical effects.
There have been few studies toward the understanding of the anomaly so far, aside from the exploration of coherent 
coupling usually related to nonlinear optical effects of four-wave mixing due to pumping and probing radiation.~\cite{lebedev}
Transient Fano resonance observed in lightly $n$-doped Si is a vestige of the quantum-mechanical effects;~\cite{hase1} though not observed in polar-semiconductors such as GaAs and GaP.~\cite{Ishioka}
The manifestation of this effect is understood based on the polaronic-quasiparticle (PQ) model, where photoexcited carriers are dressed in LO phonons by means of a strong coupling between these two particles in the ETR.~\cite{watanabe1}
A classical Fano oscillator model derived from the Fano-Anderson Hamiltonian~\cite{mahan} is also applied for this experimental result.~\cite{riffe2}
Further, based on the above PQ model, the following two effects are revealed.~\cite{watanabe2}
One is a different quantum-mechanical effect of transient plasmon-LO-phonon resonance, that is energetically resonant interaction of the plasmon with the LO phonon via dynamically
screened Coulomb interaction, which causes irregular oscillatory patterns of the CP time signals in the ETR and vanishes out of the ETR.~\cite{watanabe2}
The other is the effect of Rabi flopping with respect to pulse area of a pump laser; this effect is reflected on both of the amplitude and initial phase of the time signals in the classical region.

In this paper, we delve deeper into the issue of the transient plasmon-LO-phonon resonance.
The aim of it is to examine how the formation of such resonance affects
the degree of asymmetry of the spectral profiles associated with the CP time signals by means of both theory and experiment.
The spectral asymmetry is still one of the significant subjects in the study of the CP, and the transient Fano resonance mentioned above has been explored thus far exclusively on the view of the understanding of this subject.~\cite{hase1,watanabe1}
Here, based on the adiabatic picture with respect to time that is incorporated in the PQ model, the transient dynamics of the CP generation is described in terms of adiabatic energy curves of particles participating in this dynamics.
A particular attention is paid to the curve-crossing behavior between the plasmon and the LO-phonon, and the origin of the discernible asymmetry is analyzed by virtue of the adiabatic two-state models,~\cite{nikitin} namely, the Landau-Zener (LZ) model and the Rosen-Zener (RZ) model, as is mentioned in more detail later.
Also, the effect of optical nutation~\cite{meystre} relevant to the above-mentioned Rabi flopping on the spectral asymmetry is taken into consideration.

This paper is organized as follows. 
The theoretical framework and the experimental setup are given in Sec.~\ref{sec2} and Sec.~\ref{sec3}, respectively. 
The results and discussion are given in Sec.~\ref{sec4}, followed by the conclusions in Sec.~\ref{sec5}.
Hereafter, atomic units (a.u.) are used throughout unless otherwise stated.

\section{THEORY}
\label{sec2}

The theoretical framework of the PQ model is surveyed.
The more detail of it is already described in the previous papers.~\cite{watanabe1, watanabe2}
In short, the PQ model is a fully-quantum-mechanical model for the CP generation dynamics available for both of polar and non-polar semiconductors on an equal footing. 
The PQ operator is defined as composed of two kinds of operators; one is a quasiboson operator, given by a linear combination of a set of pairs of electron operators, and the other is an LO-phonon operator. Based on this PQ model, one can track the CP dynamics in the ETR. 
The total Hamiltonian of concern is composed of an electron Hamiltonian, an interaction between an electron and an external pump-pulse laser, an LO phonon Hamiltonian, and an electron-LO-phonon interaction.
Here, a Coulomb potential interaction is incorporated with the electron Hamiltonian described by a two-band model including conduction ($c$) and valence ($v$) bands.
The interaction of the laser with electron in band $b$ at time $t$ is given by
\begin{equation}
\Omega_{b\bar{b}}(t)=\Omega_{0b\bar{b}} f(t) \cos \omega_0 t,
\label{Omega(t)}
\end{equation}
where $\Omega_{0b\bar{b}}$ represents the Rabi frequency~\cite{haug} given by the product of peak electric-field strength of the laser pulse and the electric dipole moment between $\Gamma$-points of $c$ and $v$ bands. 
Here $b$ is either $c$ or $v$ with $\bar{b} \neq b$, 
$f(t)$ represents the pulse envelope function given by the Gaussian function with temporal width (the full width at half maximum) $\tau_L$, and $\omega_0$ is the center frequency of the external laser-field.
Hereafter, it is understood that creation and annihilation operators of electron with Bloch momentum $\boldsymbol{k}$ in band $b$ are represented by $a^\dagger_{b\boldsymbol{k}}$ and $a_{b\boldsymbol{k}}$, respectively, and 
creation and annihilation operators of LO phonon with energy dispersion $\omega_{\boldsymbol{q}}$ at Bloch momentum $\boldsymbol{q}$ are represented by $c_{\boldsymbol{q}}^\dagger$ and $c_{\boldsymbol{q}}$, respectively.

We consider the time-evolution of a composite operator defined by $A_{\boldsymbol{q}}^\dagger(\boldsymbol{k}bb') = a^\dagger_{b\boldsymbol{k+q}} a_{b^\prime\boldsymbol{k}}$ on the basis of adiabatic approximation with respect to $t$.~\cite{watanabe1, watanabe2}
$A_{\boldsymbol{q}}^\dagger(\boldsymbol{k}bb')$ represents a carrier density matrix for the transition from $b'$-band to $b$-band with an anisotropic momentum distribution determined by the transferred momentum $\boldsymbol{q}$; this is quite small, but finite ($\boldsymbol{q} \neq 0$). 
Here, the rotating wave approximation~\cite{haug} is employed in order to remove high-frequency contributions from the equations of motion, and thus
$A_{\boldsymbol{q}}^\dagger(\boldsymbol{k}bb')$ is replaced by
$\bar{A}_{\boldsymbol{q}}^\dagger(\boldsymbol{k}bb')=
A_{\boldsymbol{q}}^\dagger(\boldsymbol{k}bb')e^{-i \bar{\omega}_{bb'}t}$,
where $\bar{\omega}_{cv}=\omega_0$, $\bar{\omega}_{vc}=-\omega_0$, and $\bar{\omega}_{bb}=0$.
Further a creation operator of a collective excitation mode --- a plasmon --- $B^\dagger_{\boldsymbol{q}}$ with the plasma frequency $\omega_{\boldsymbol{q}pl}$ is introduced by a linear combination of the intraband density matrices $A_{\boldsymbol{q}}^\dagger(\boldsymbol{k}bb)$'s.~\cite{watanabe2}
As regards the single-particle excitation mode, just the interband contributions from $A_{\boldsymbol{q}}^\dagger(\boldsymbol{k}b\bar{b})$'s are retaimed, and the intraband contributions from $A_{\boldsymbol{q}}^\dagger(\boldsymbol{k}bb)$'s are neglected because these vanish in the long wave-length limit ($|\boldsymbol{q}|\rightarrow 0$).~\cite{watanabe2, haug}

The equations of motion of $c_{\boldsymbol{q}}^\dagger$, $\bar{A}_{\boldsymbol{q}}^\dagger(\boldsymbol{k}b\bar{b})$, and $B^\dagger_{\boldsymbol{q}}$ are provided in a matrix form of~\cite{watanabe2}
\begin{equation}
-i {d \over dt}\left[
c_{\boldsymbol{q}}^\dagger , \bar{A}_{\boldsymbol{q}}^\dagger (\boldsymbol{k}b\bar{b}) , B_{\boldsymbol{q}}^\dagger
\right]
=
\left[
c_{\boldsymbol{q}}^\dagger, \bar{A}_{\boldsymbol{q}}^\dagger (\boldsymbol{k}b\bar{b}), B_{\boldsymbol{q}}^\dagger
\right] \bar{Z}_{\boldsymbol{q}}. 
\label{HeisenbergcAB}
\end{equation}
Here, $\bar{Z}_{\boldsymbol{q}}=\{ \bar{Z}_{\boldsymbol{qjj'}}\}$ represents a non-Hermitian matrix, the explicit form of which is given in Ref.~\onlinecite{watanabe2}, and
hereafter, it is understood that $j, j'=\{ph, (\boldsymbol{k}b\bar{b}), pl \}$ with the LO-phonon mode $ph$, the single-particle excitation mode $(\boldsymbol{k}b\bar{b})$, and the plasmon mode $pl$.
In Eq.~(\ref{HeisenbergcAB}), phenomenological damping constants for all of these modes are suppressed.
In addition, 
the argument of $t$ in $c_{\boldsymbol{q}}^\dagger, \bar{A}_{\boldsymbol{q}}^\dagger (\boldsymbol{k}b\bar{b}), B_{\boldsymbol{q}}^\dagger$, and $\bar{Z}_{\boldsymbol{q}}$ is omitted just for the sake of simplicity, and hereafter this convention is used unless otherwise stated.
Now, we introduce the PQ operator as follows:
\begin{equation}
P_{\boldsymbol{q}j}^\dagger = c_{\boldsymbol{q}}^\dagger V_{\boldsymbol{q}ph j}^R
+ B_{\boldsymbol{q}}^\dagger V_{\boldsymbol{q}pl j}^R
+ \sum_{\boldsymbol{k}b} \bar{A}_{\boldsymbol{q}}^\dagger (\boldsymbol{k}b\bar{b}) V_{\boldsymbol{q}(\boldsymbol{k}b\bar{b}) j}^R,
\label{PQoperator}
\end{equation}
where left and right eigenvalue problems~\cite{Moiseyev} of $\bar{Z}_{\boldsymbol{q}}$ as $V_{\boldsymbol{q}j}^{L\dagger} \bar{Z}_{\boldsymbol{q}}= {E}_{\boldsymbol{q}j}V_{\boldsymbol{q}j}^{L\dagger}$ and $\bar{Z}_{\boldsymbol{q}}V_{\boldsymbol{q}j}^{R}=V_{\boldsymbol{q}j}^{R} {E}_{\boldsymbol{q}j}$, respectively, are solved in an adiabatic sense to obtain the $j$th eigenvalue ${E}_{\boldsymbol{q}j}$ and the corresponding biorthogonal set of eigenvectors $\{ V_{\boldsymbol{q}j}^{L}, V_{\boldsymbol{q}j}^{R} \}$.
The time-evolution of $P_{\boldsymbol{q}j}^\dagger$ is obtained by solving the associated Heisenberg equation, where a non-adiabatic coupling between $j$th and $j'$th modes are assumed negligibly small in the practical calculations.~\cite{watanabe2}

In the linear response theory, the retarded phonon Green function $D_{\boldsymbol{q}}^R(t,t')$ indicates an induced charge density of ionic cores probed at time $t'$ by a weak external potential with a delta-function form $\delta(t')$.~\cite{schafer}
This is provided by $D_{\boldsymbol{q}}^R(t,t')
=\bar{D}_{\boldsymbol{q}}^R(t,t') + \left[ \bar{D}_{\boldsymbol{-q}}^R(t,t') \right]^*$ in terms of the PQ operator with the relation of $c_{\boldsymbol{q}}^\dagger = \sum_{j} P_{\boldsymbol{q}j}^\dagger V_{\boldsymbol{q}j, ph}^{L\dagger}$ as follows:
\begin{eqnarray}
\bar{D}_{\boldsymbol{q}}^R(t,t')
&=&
-i \left \langle \left [c_{\boldsymbol{q}}(t), c_{\boldsymbol{q}}^\dagger (t') \right] \right \rangle
\theta (t-t') \nonumber \\
&=&
-i \sum_{jj'} V_{\boldsymbol{q}ph, j}^L(t) 
\left \langle \left[
P_{\boldsymbol{q}j} (t), P_{\boldsymbol{q}j'}^\dagger (t') 
\right] \right \rangle \nonumber \\
&& \times V_{\boldsymbol{q}j', ph}^{L\dagger}(t') 
\theta (t-t')
\label{DRbar1}
\end{eqnarray}
and $ \bar{D}_{\boldsymbol{-q}}^R(t,t')= \bar{D}_{\boldsymbol{q}}^R(t,t')$,
where the ground-state expectation value is taken in $\langle \cdots \rangle$.
The induced charge density attributed to the CP generation is provided by
\begin{equation}
Q_{\boldsymbol{q}}(\tau)\equiv D_{\boldsymbol{q}}^R(\tau+t',t')-D_{\boldsymbol{q}}^{R(0)}(\tau+t',t')
\label{Q0}
\end{equation}
with $\tau=t-t' \ge 0$,
and it is considered that $Q_{\boldsymbol{q}}(\tau)$ corresponds to a displacement function of the CP at time $\tau$ aside from an unimportant proportional constant. 
Here, the contribution of the incoherent phonon signal is subtracted, where this is given by the free phonon Green function $D_{\boldsymbol{q}}^{R(0)}(\tau+t',t')$ without the pump laser, and hereafter the probing time of $t'=0$ is exclusively concerned.
The oscillatory pattern of the CP is extracted by representing $Q_{\boldsymbol{q}}(\tau)$ as
\begin{equation}
Q_{\boldsymbol{q}}(\tau) = A_{\boldsymbol{q}}(\tau) \cos \left[\omega_{\boldsymbol{q}}\tau+\Theta_{\boldsymbol{q}}(\tau) \right],
\label{Q}
\end{equation}
where $\Theta_{\boldsymbol{q}}(\tau) $ and $A_{\boldsymbol{q}}(\tau)$ represent a renormalized phase modulus $\pi$ and a transitory amplitude at $\tau$, respectively.
The Fourier transform (FT) of $Q_{\boldsymbol{q}}(\tau)$ is provided by
\begin{equation}
\tilde{Q}_{\boldsymbol{q}}(\omega) = \int_{0}^\infty e^{-i \omega \tau} Q_{\boldsymbol{q}}(\tau) d\tau,
\label{Qtilde}
\end{equation}
and the associated power spectrum $S_{\boldsymbol{q}}(\omega)$ is given by
\begin{equation}
S_{\boldsymbol{q}}(\omega) \propto |\tilde{Q}_{\boldsymbol{q}}(\omega)|^2.
\label{Sq}
\end{equation}

In the long-time limit of $\tau$ ensuring $\tau \gg T_{\boldsymbol{q}} \equiv 2\pi / \omega_{\boldsymbol{q}}$, $\Theta_{\boldsymbol{q}}(\tau)$ and $A_{\boldsymbol{q}}(\tau)$ become $\theta_{\boldsymbol{q}}$ and $A^0_{\boldsymbol{q}} \:e^{-\tau /T_{\boldsymbol{q}ph}}$, respectively, where $\theta_{\boldsymbol{q}}$ and $A^0_{\boldsymbol{q}}$ are constants and $T_{\boldsymbol{q}ph}$ represents a relaxation time constant attributed to phonon anharmonicity. 
In this time region, Eq. (\ref{DRbar1}) becomes
\begin{equation}
\bar{D}_{\boldsymbol{q}}^R(\tau,0)
=
-i e^{-i\omega_{\boldsymbol{q}}\tau} \xi_{\boldsymbol{q}}(\tau,0) 
V_{\boldsymbol{q}ph,ph}^{L\dagger}(0) \:\theta(\tau),
\label{DRbar3}
\end{equation}
where the relations of $V_{\boldsymbol{q}ph,j}^{R\dagger}(-\infty)= \delta_{ph,j}$, 
$V_{\boldsymbol{q}ph,ph}^L(\tau)\approx1$,
$\left [ c_{\boldsymbol{q}} (-\infty), c_{\boldsymbol{q}}^\dagger (-\infty) \right]=1$, and $\left [ c_{\boldsymbol{q}}, B_{\boldsymbol{q}}^\dagger \right] =\left[ c_{\boldsymbol{q}}, A_{\boldsymbol{q}}^\dagger(\boldsymbol{k}b\bar{b}) \right] =0$ are used.
In fact, $\xi_{\boldsymbol{q}}(\tau,0) $ given by
\begin{eqnarray}
&&\xi_{\boldsymbol{q}}(\tau,0) \nonumber \\
&& \approx
\exp \left [
-\int_{-\infty}^\tau dt'' {\rm Im} E_{\boldsymbol{q}ph}(t'') - \int_{-\infty}^{0} dt'' {\rm Im} E_{\boldsymbol{q}ph}(t'')
\right ] \nonumber \\
&& \ \times \exp \left [
-i \int_{0}^\tau dt'' \left \{ {\rm Re} E_{\boldsymbol{q}ph}(t'') - \omega_{\boldsymbol{q}} \right \}
\right ]
\label{xi}
\end{eqnarray}
is approximately unity.
Thus, $\theta_{\boldsymbol{q}}$ and $A^0_{\boldsymbol{q}}$ are provided as
\begin{equation}
\theta_{\boldsymbol{q}} = {\pi \over 2} -{\rm arg}
\left[\delta V_{\boldsymbol{q}ph} \right]
\label{theta}
\end{equation}
modulus $\pi$, and
\begin{equation}
A^0_{\boldsymbol{q}} = \left|\delta V_{\boldsymbol{q}ph} \right|,
\label{A0}
\end{equation}
respectively, where
\begin{equation}
\delta V_{\boldsymbol{q}ph}=V_{\boldsymbol{q}ph,ph}^{L\dagger}(0) -1.
\label{deltaV}
\end{equation}

Numerical calculations for undoped Si are conducted.
Material parameters employed here are given in Ref.~\onlinecite{watanabe1}
with $\tau_L=10$ fs, 
$T_{\boldsymbol{q}}=66$ fs equivalent to $\omega_{\boldsymbol{q}}=63$ meV = 15.2 THz, and $T_{\boldsymbol{q}ph}=2$ ps.

\section{EXPERIMENTS}
\label{sec3}
The anisotropic transient reflectivity of $n$-doped ($\sim$ 1$\times$10$^{15}$cm$^{-3}$) Si(001) sample is measured in air at 
295 K\cite{hase1} by the electro-optic (EO) sampling technique.\cite{Pfeifer92EO} Nearly collinear, pump and probe beams --- with center frequency $\omega_p =$ 2.989 - 3.180 eV (390 - 415 nm) ---  are overlapped at a 7.2$\times$10$^{-7}$cm$^{2}$ spot on the sample. The maximum 
average pump power of 18 mW generates $N$ $\approx$ 3.0$\times$10$^{19}$cm$^{-3}$ carriers estimated from the absorption coefficient   $\alpha$ = 8.0$\times$10$^{4}$cm$^{-1}$ at 
397 nm.\cite{Hulthen75} This is 10 times less than the critical density for screening of the carrier-phonon interaction.\cite{Sjodin98} 
After reflecting from the sample, the probe is analyzed into polarization components parallel and perpendicular to that of the 
pump and each is detected with a photodiode. The resulting photocurrents are subtracted and their difference ($\Delta R_{eo}/R$ = ($\Delta R_{\parallel} - \Delta R_{\perp})/R$) was 
recorded versus the pump-probe delay. The delay is scanned over 8 ps at 20 Hz frequency.\cite{hase2} 

\section{RESULTS AND DISCUSSION}
\label{sec4}

\begin{figure}[tb]
\begin{center}
\includegraphics[width=7.5cm,clip]{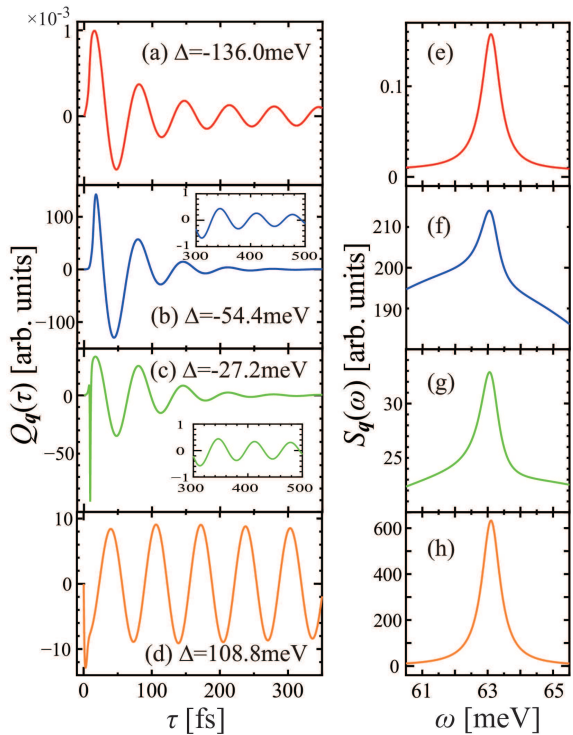}
\caption{The calculated results of the oscillatory pattern $Q_{\boldsymbol{q}}(\tau)$ as a function of $\tau$ (in the unit of fs) and the associated power spectrum $S_{\boldsymbol{q}}(\omega)$ as a function of $\omega$ (in the unit of meV) for the detuning $\Delta = $ (a) and (e) $-136$ meV, (b) and (f) $-54.4$ meV, (c) and (g) $-27.2$ meV, and (d) and (h) 108.8 meV. The insets in Figs. (b) and (c) show $Q_{\boldsymbol{q}}(\tau)$ for 300 fs $\leq \tau \leq$ 500 fs.}
\label{fig1}
\end{center}
\end{figure}

Figure~\ref{fig1} shows the calculated results of $Q_{\boldsymbol{q}}(\tau)$ as a function of $\tau$ and the associated $S_{\boldsymbol{q}}(\omega)$ for $\Delta =-136$ meV, $-54.4$ meV, $-27.2$ meV, and 108.8 meV, respectively with $\Omega_{0cv}=$108.8 meV.
Here, $\Delta$ represents the detuning defined as the difference of the center frequency of a pump-laser pulse $\omega_0$ from the direct band gap energy at $\Gamma$ point $E_g$, that is,
$\Delta = \omega_0 - E_g$.
$Q_{\boldsymbol{q}}(\tau)$ in Fig.~\ref{fig1}(a) looks almost sinusoidal. 
The spectrum $S_{\boldsymbol{q}}(\omega)$ in Fig.~\ref{fig1}(e) peaks at $\omega = \omega_{\boldsymbol{q}}$ with a symmetric profile.
It is seen that in Figs.~\ref{fig1}(b) and~\ref{fig1}(c), the amplitudes of $Q_{\boldsymbol{q}}(\tau)$ in the ETR ($\tau \lesssim 100$ fs) are much larger than those in the classical region ($\tau \gtrsim 100$ fs), and the spectral profiles are found asymmetric in $S_{\boldsymbol{q}}(\omega)$ seen in Fig.~\ref{fig1}(f) and \ref{fig1}(g).
In contrast, in Fig.~\ref{fig1}(d), $Q_{\boldsymbol{q}}(\tau)$ oscillates almost in a sinusoidal manner with a constant amplitude to show a symmetric spectral profile as seen in Fig.~\ref{fig1}(h).
The irregularity with the enhancement of amplitudes in the ETR shown in Figs.~\ref{fig1}(b) and~\ref{fig1}(c) is attributed to the almost energetically resonant interaction between both modes of plasmon and LO-phonon, as discussed in Ref.~\onlinecite{watanabe2}, so is 
the associated spectral asymmetry.

\begin{figure}[tb]
\begin{center}
\includegraphics[width=7.5cm,clip]{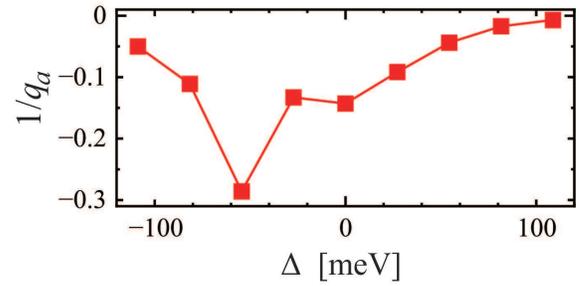}
\caption{The inverse of the calculated asymmetry parameter $1/q_a$ for the power spectrum $S_{\boldsymbol{q}}(\omega)$ as a function of $\Delta$ (in the unit of meV).}
\label{fig2}
\end{center}
\end{figure}

The asymmetry observed in $S_{\boldsymbol{q}}(\omega)$ of Fig.~\ref{fig1} is evaluated by fitting these spectra in the vicinity of each peak to Fano's spectral formula,~\cite{fano} \begin{eqnarray}
I(\omega) = C\frac{(q_a + \varepsilon)^{2}}{1 + \varepsilon^{2}} + const.,
\label{fanolineshape}
\end{eqnarray}   
where $\varepsilon = (\omega - \omega_{r} - \Delta \omega)/\Gamma$, $C$ is the amplitude, $q_a$ is the asymmetry parameter, $\omega_{r}$ is the unperturbed frequency, 
and the frequency shift $\Delta \omega$ and the broadening parameter $\Gamma$ are the carrier density dependent real and imaginary parts of the LO phonon self-energy associated 
with the interaction with photogenerated carriers.~\cite{hase1,Cerderia73}
Figure~\ref{fig2} shows the inverse of the asymmetry parameter $1/q_a$ having a negative value as a function of $\Delta$ with $\Omega_{0cv}=$108.8 meV;
where a spectrum $S_{\boldsymbol{q}}(\omega)$ becomes symmetric with the decrease of $1/|q_a|$, while this becomes asymmetric with the increase of it aside from $1/|q_a| \gg1$ showing a (symmetric ) spectral dip.
At $\Delta < -100$ meV, $1/|q_a|$ is reduced.
As $\Delta$ increases, $1/|q_a|$ becomes larger, because the plasmon-phonon resonance likely occurs in the ETR. 
Within our calculations, the profile is the most asymmetric at $\Delta=-54.4$ meV with $1/q_a=-0.286$.
As $\Delta$ further increases, $1/q_a$ approaches zero again to show symmetric profiles.

\begin{figure}[tb]
\begin{center}
\includegraphics[width=8.6cm,clip]{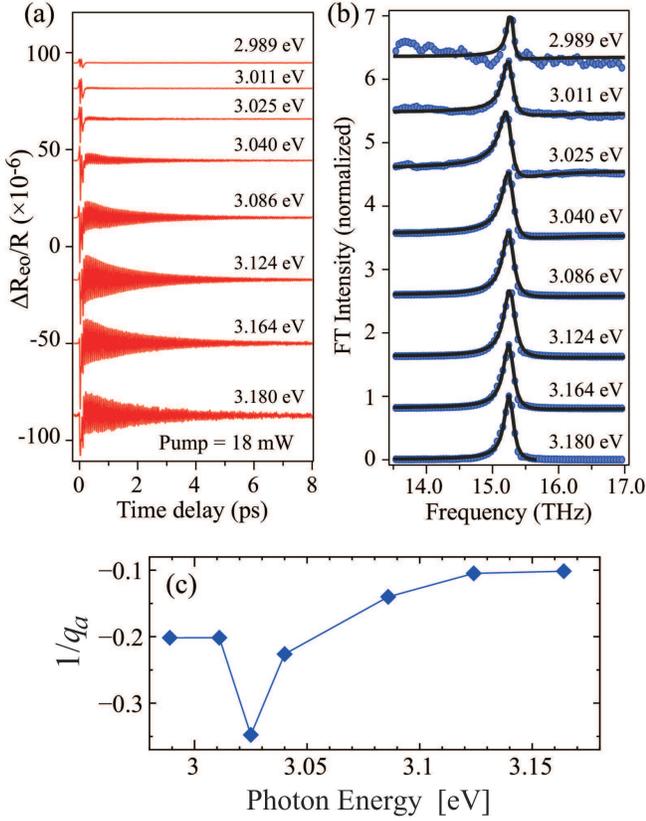}
\caption{The experimental results in lightly $n$-doped Si for different photon energy $\omega_p$ of (a) the transient electro-optic reflectivity as a function of time delay (in the unit of ps), (b) the associated FT power spectra of the reflectivity signals as function of frequency (in the unit of THz).
The black solid line represent a fit using Eq. (\ref{fanolineshape}).
(c) The inverse of the asymmetric parameter $1/q_a$ for the power spectrum as a function of $\omega_p$ (in the unit of eV).}
\label{fig3}
\end{center}
\end{figure}

Figure~\ref{fig3}(a) shows the experimental results of transient EO reflectivity as a function of time delay for different photon energy $\omega_p$, and the FTs of these oscillatory signals are shown in Fig.~\ref{fig3}(b). 
For the 2.989 eV excitation, aperiodic electronic response near the zero delay dominates the signal. For higher energies, in addition, there appears a coherent oscillation with a period 
of $\sim$ 66 fs that persists for $\sim$ 8 ps due to the $|\boldsymbol{q}| \approx 0$ coherent LO phonons.\cite{hase1,Cerderia73} The phonon amplitude monotonically increases with increasing the photon energy, reaching a maximum of $\Delta R_{eo}/R$ $\sim$ 2.5$\times$10$^{-5}$ at the high-energy limit of the tuning range, where it is comparable to the electronic response. 
As with the spontaneous Raman spectra, the LO phonon signal is enhanced by resonance with the direct band gap of Si (attributed to two nearly overlapping transitions E'$_{0}$ (3.320 eV) at the $\Gamma$ point and the more intense E$_{1}$ (3.396 eV) for a range of momenta along $\Lambda$\cite{Renucci75}). 
The FT spectra show a pronounced asymmetry, which we fit to a Fano lineshape
given by Eq.~(\ref{fanolineshape}).~\cite{fano,Cerderia73}

Further, Fig.~\ref{fig3}(c) shows the inverse of the asymmetry parameter $1/q_a$ as a function of $\omega_p$, which is extracted from the signals of Fig~\ref{fig3}(b) in the same manner as that of Fig~\ref{fig2}.
The intensity of the pulse laser irradiated here corresponds to $\Omega_{0cv}=$110 meV almost equal to that used in the theoretical calculations for Figs.~\ref{fig1} and \ref{fig2}.
In actual experiments accompanying high-density carrier excitation, a primitive bandgap $E_g\approx 3.3$ meV is likely modified to $\bar{E}_g$ due to the effect of bandgap renormalization.~\cite{hase2}
Accordingly, it is considered that the photon energy $\omega_p$ necessary for the carrier excitation is reduced to some extent from $\omega_0$, that is,
$\omega_p=\omega_0-\Delta_g$ with the magnitude of the bandgap renormalization as $\Delta_g \equiv E_g-\bar{E}_g$.
Thus, the detuning is represented by $\Delta=\omega_p-\bar{E}_g$.
As shown in Figs.~\ref{fig3}(b) and \ref{fig3}(c), for $\omega_p \lesssim 3.05$ eV, the spectral profiles are asymmetric, and the profile is the most asymmetric at $\omega_p =3.025$ eV with $1/q_a=-0.348$, where the amplitudes of the time signals in the ETR are much larger than those in the classical region.
On the other hand, for $\omega_p \gtrsim 3.05$ eV, as $\omega_p$ increases, the profile becomes symmetric.
These results are in agreement with the calculated results shown in Figs.~\ref{fig1}, and \ref{fig2}, if the estimate of $\Delta_g$ is considered to be roughly 200 meV.
Actually, the photoexcited carrier density is $2.0 \times 10^{19}$ cm$^{-3}$ for $\Delta=54.4$ meV in the present calculations, leading to $\omega_0\approx 3.35$ eV.
On the other hand, the photoexcited carrier density is $3.0 \times 10^{19}$ cm$^{-3}$ for $\omega_p=3.12$ meV in the experiment cited in Ref.~\onlinecite{hase2}.
Because both of the photoexcited carrier densities are almost equal,
we obtain $\Delta_g \approx 230$ meV, which is the same order of magnitude as that calculated by a first-principle technique and the GW approximation for the self-energy operator.~\cite{oschlies}

\begin{figure}[tb]
\begin{center}
\includegraphics[width=7.0cm,clip]{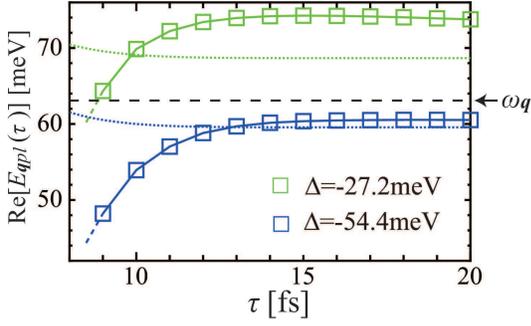}
\caption{The real part of the adiabatic eigenvalue of the plasmon mode Re[$E_{\boldsymbol{q}pl}(\tau)$] (in the unit of meV) as a function of $\tau$ (in the unit of fs) for $\Delta =-54.4$ meV (blue square) and $-27.2$ meV (green square). The blue and green dot lines represent the associated plasma frequencies $\omega_{\boldsymbol{q}pl}(\tau)$, and the black dash line represents the bare phonon energy $\omega _{\boldsymbol{q}}$.}
\label{fig4}
\end{center}
\end{figure}

Next, with the goal of 
deepening the understanding of the asymmetry observed in Fig.~\ref{fig2}, first, we examine the detail of the pronounced behavior of the CP signals shown in Figs.~\ref{fig1}(b) and \ref{fig1}(c) in the time region of $\tau \lesssim 20$ fs much smaller than the period $T_{\boldsymbol{q}}$.
Figure~\ref{fig4} shows the real parts of adiabatic energy curves of plasmon mode $E_{\boldsymbol{q}pl}(\tau)$ with respect to $\tau$ for $\Delta = -54.4$ and $-27.2$ meV with $\Omega_{0cv}=$108.8 meV.
$E_{\boldsymbol{q}pl}(\tau)$'s are renormalized by the electron-laser interaction, the Hartree-Fock interaction, and the electron-LO-phonon interaction to lead to the
deviation from an unrenormalized plasmon frequency $\omega_{\boldsymbol{q}pl}(\tau)$ to some extent.
In particular, this stands out in the region of $\tau \lesssim 10$ fs.
Actually, at $\tau =9$ fs, despite $ \tau > \tau_L/2$, 
Re[$E_{\boldsymbol{q}pl}(\tau)$]'s differ from $\omega_{\boldsymbol{q}pl}(\tau)$ by roughly 10 and 5 meV for $\Delta =-54.4$ and -27.2 meV, respectively, due to the renormalization of still non-negligible electron-laser interaction in the tail region of the pulse, where $\Omega_{cv}(\tau)=11.5$ meV.
As time passes, $\Omega_{cv}(\tau)$ vanishes, and Re$[E_{\boldsymbol{q}pl}(\tau)]$ becomes closer to $\omega_{\boldsymbol{q}pl}(\tau)$ aside from the energy difference between these two energy-values due to the Hartree-Fock interaction and the electron-LO-phonon interaction that are still retained in the ETR, as seen in Fig.~\ref{fig4}.

Concerning another adiabatic energy of LO-phonon mode $E_{\boldsymbol{q}ph}(\tau)$, the real part of it, Re$[E_{\boldsymbol{q}ph}(\tau)]$, is almost equal to $\omega_{\boldsymbol{q}}$ within a couple of meV due to the renormalization of the electron-LO-phonon interaction;~\cite{watanabe2}
thus, for the sake of simplicity, hereafter, let $E_{\boldsymbol{q}ph}(\tau)$ be represented 
by $\omega_{\boldsymbol{q}}$ in Fig.~\ref{fig4}.
The way of interaction between the plasmon and LO-phonon modes
can be understood in terms of the following two-state models in the adiabatic picture as
the LZ model and the RZ model.~\cite{nikitin}
In the LZ model, two adiabatic-energy curves tend to swerve sharply around the local time $\tau=\tau_{LZ}$, at which both are closest to each other forming anti-crossing, while in the RZ model, these curves evolve in a parallel manner over time $\tau$ after $\tau_{RZ}$ with almost energetically degeneracy.
It is seen in Fig.~\ref{fig4} that for $\Delta = -27.2$ meV, Re$[E_{\boldsymbol{q}pl}(\tau)]$ traverses Re$[E_{\boldsymbol{q}ph}(\tau)]$ around $\tau_{LZ}\equiv 9$ fs, while  for $\Delta = -27.2$ and -54.4 meV, Re$[E_{\boldsymbol{q}pl}(\tau)]$ becomes constant after  $\tau_{RZ}\equiv 12$ fs to form the RZ coupling in the ETR.
Let the energy difference of Re$[E_{\boldsymbol{q}pl}(\tau)]$ from Re$[E_{\boldsymbol{q}ph}(\tau)]$ in the ETR be represented by $\delta E_{\boldsymbol{q}}$, that is, $\delta E_{\boldsymbol{q}} \equiv {\rm Re}[E_{\boldsymbol{q}pl}(\tau)]-{\rm Re}[E_{\boldsymbol{q}ph}(\tau)]$; where $\delta E_{\boldsymbol{q}}$ is considered almost constant in $\tau >\tau_{RZ}$.
It is evident that with the decrease of $|\delta E_{\boldsymbol{q}}|$, the effect of the RZ coupling becomes more significant.
Actually,  $\Delta$'s in  Fig~\ref{fig1} are put  in an increasing order of  $|\delta E_{\boldsymbol{q}}|$   as  -54.4, -27.2, -136.0, and 108.8 meV.
Incidentally, for $\Delta =108.8$ meV, an anticrossing due to the LZ coupling likely occurs just in the region of $\tau\lesssim \tau_L/2$, though the overall oscillatory pattern is little affected by this.

\begin{figure}[tb]
\begin{center}
\includegraphics[width=7.0cm,clip]{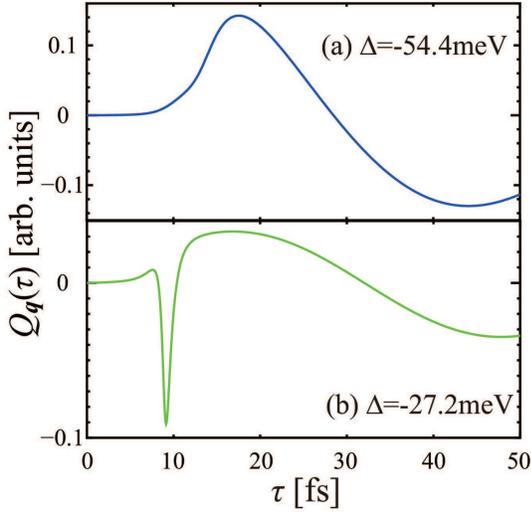}
\caption{The enlarged view of Figs.~\ref{fig1}(b) and \ref{fig1}(c) in the limited temporal region of $\tau$ for $\Delta=$ (a) -54.4 meV and (b) -27.2 meV, respectively.}
\label{fig5}
\end{center}
\end{figure}

The interaction between the two modes of plasmon and LO-phonon causes abrupt changes of $A_{\boldsymbol{q}}(\tau)$ and $\Theta_{\boldsymbol{q}}(\tau)$ in the oscillatory pattern of $Q_{\boldsymbol{q}}(\tau)$.~\cite{watanabe2}
Figure~\ref{fig5} is the enlarged view of Figs.~\ref{fig1}(b) and \ref{fig1}(c) in the limited region of $\tau \le 50$ fs to show the detailed behavior of $Q_{\boldsymbol{q}}$ for $\Delta=-54.4$ and -27.2 meV, respectively.
In Fig.~\ref{fig5}(a), $A_{\boldsymbol{q}}(\tau)$ becomes pronouncedly enhanced around $\tau_{RZ}$.
This consists with the formation of the RZ coupling between the plasmon and LO-phonon modes, as is seen in Fig.~\ref{fig4}.
Concerning the case of Fig.~\ref{fig5}(b), both of $A_{\boldsymbol{q}}(\tau)$ and $\Theta_{\boldsymbol{q}}(\tau)$ change sharply just in the vicinity of $\tau_{LZ}$,
followed by the enhancement of $A_{\boldsymbol{q}}(\tau)$ in the region of $\tau \gtrsim \tau_{RZ}$.
Also, this consists of the instantaneous formation of the LZ coupling between the two modes, followed by the RZ coupling in the rest of the ETR, as is seen in Fig.~\ref{fig4}.
In addition, for $\Delta = -136.0$ and 108.8 meV, $Q_{\boldsymbol{q}}$'s oscillate in a more sinusoidal manner than those for $\Delta = -54.4$ and -27.2 meV, as is seen in Figs.~\ref{fig1}(a) and \ref{fig1}(d), respectively.

\begin{figure}[tb]
\begin{center}
\includegraphics[width=7.0cm,clip]{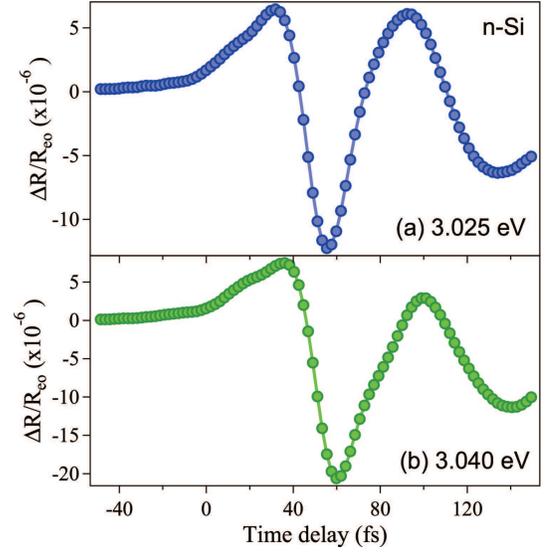}
\caption{The enlarged view of Fig.~\ref{fig3}(a) in the limited temporal region for $\omega_p=$ (a) 3.025 eV and (b) 3.040 eV.}
\label{fig6}
\end{center}
\end{figure}

Figure~\ref{fig6} is the enlarged view of Fig.~\ref{fig3}(a) for $\omega_p=3.025$ and 3.040 eV in the limited region.
It is seen that transient electro-optic reflectivity signals oscillate with anomalously larger amplitudes and more varying phases in time, compared with those in the classical region.
These oscillatory patterns are in overall agreement with those in Fig.~\ref{fig5},
aside from the rapid change of $\Theta_{\boldsymbol{q}}(\tau)$ observed in Fig.~\ref{fig5}(b).
It is remarked that such a rapid change caused by the LZ coupling would be possibly modified by the non-adiabatic correction, because this correction is considered still effective to some extent in the tail region of the pulse in $\tau \lesssim 10$ fs; though
 neglected here, as mentioned in Sec.~\ref{sec2}.
To be specific, a diagonal part of the correction would modify the adiabatic energy Re$[E_{\boldsymbol{q}pl}(\tau)]$, while the off-diagonal parts of it, especially, providing the interaction of the plasmon mode with other modes would 
blur the location of the crossing around $\tau_{LZ}$.

Now, more detailed discussion is made on the asymmetry observed in Figs.~\ref{fig2} and \ref{fig3}(c).
To do this, the region of $\tau$ for the CP signals shown in Fig.~\ref{fig1}(a) is divided into two regions inside and outside the ETR, namely, $\tau \lesssim 100$ fs and $\tau \gtrsim 100$ fs, respectively.
Accordingly, the power spectrum $S_{\boldsymbol{q}}(\omega)$ is cast into the sum of the contributions from the inner region $S^{(in)}_{\boldsymbol{q}}(\omega)$ and the outer region $S^{(out)}_{\boldsymbol{q}}(\omega)$, that is,
\begin{equation}
S_{\boldsymbol{q}}(\omega)=S^{(in)}_{\boldsymbol{q}}(\omega)+S^{(out)}_{\boldsymbol{q}}(\omega).
\label{S}
\end{equation}
It is obvious that $S^{(out)}_{\boldsymbol{q}}(\omega)$ shows the Lorentzian profile with the peak frequency of $\omega_{\boldsymbol{q}}$ and the full-width at half maximum of $1/T_{\boldsymbol{q}ph}$.
By contrast the spectral profile of $S^{(in)}_{\boldsymbol{q}}(\omega)$ is straightforwardly 
affected by the oscillatory pattern of $Q_{\boldsymbol{q}}(\tau)$ in the ETR.
The profile of $S^{(in)}_{\boldsymbol{q}}(\omega)$ is more or less deformed from the Lorentzian shape due to the aperiodic oscillatory-patterns of $Q_{\boldsymbol{q}}(\tau)$, as seen in Fig.~\ref{fig1}.
Hence, $S_{\boldsymbol{q}}(\omega)$ proves asymmetric if $S^{(in)}_{\boldsymbol{q}}(\omega)$ is comparable to or dominant over $S^{(out)}_{\boldsymbol{q}}(\omega)$;
otherwise this is symmetric.
The asymmetric spectra of Figs.~\ref{fig1}(f) and \ref{fig1}(g) are the case.
It is noted that the anomaly arising from the LZ coupling shown in Fig.~\ref{fig5}(b) is less effective against $S^{(in)}_{\boldsymbol{q}}(\omega)$ than the contribution from the RZ coupling, since the former coupling is localized just in vicinity of $\tau_{LZ}$.
To be more specific, the degree of asymmetry in $S_{\boldsymbol{q}}(\omega)$ is mostly governed by the 
magnitude of the RZ coupling in the ETR that is evaluated in terms of $\delta E_{\boldsymbol{q}}$.
Actually, with the increase in $|\delta E_{\boldsymbol{q}}|$, $S_{\boldsymbol{q}}(\omega)$ is apt to be more symmetric within the scope of the calculations implemented here, as is shown in Figs.~\ref{fig2} and \ref{fig3}(c).
Incidentally, $\delta E_{\boldsymbol{q}}$ is determined by the Rabi frequency $\Omega_{0cv}$ as well as $\Delta$, because Re$[E_{\boldsymbol{q}pl}(\tau)]$ as well as $\omega_{\boldsymbol{q}pl}$ is subject to the photoexcited carrier density.

Below, we give consideration to the alteration pattern of $1/q_a$ when a pulse area represented by $S_L$ exceeds $\pi$, that is, $S_L/\pi >1$; this pulse is termed as pulse 1 hereafter.
For the sake of simplicity, $S_L$ is assumed to be evaluated as the product of the nutation frequency given by
\begin{equation}
\Omega_N = \sqrt{\Omega_{0cv}^2+\Delta^2}
\label{OmegaN}
\end{equation}
and $\tau_L$, that is, $S_L=\Omega_N\tau_L$,
where a dispersionless two-level model is adopted without the Coulomb correction to $\Omega_{0cv}$.
When the pump-pulse irradiation is completed, the total amount of photoexcited carrier density per site, denoted as $D_L$, is given by
\begin{equation}
D_L=\left({\Omega_{0cv}\over \Omega_N}\right)^2\sin^2\left({S_L\over 2}\right).
\label{DL}
\end{equation}
This expression is obtained by employing a square pulse with $f(t)=\theta(t+\tau_L/2)\theta(\tau_L/2-t)$ in Eq.~{(\ref{Omega(t)}) instead of the Gaussian pulse adopted above.
Another pulse termed as pulse 0 with pulse area $S^0_L < \pi $ is introduced, where this pulse has a nutation frequency $\Omega^0_N$ with $\Omega^0_{0cv}$ and $\Delta^0$ as Rabi frequency and detuning, respectively;
$\Omega^0_N = \sqrt{(\Omega^0_{0cv})^2+(\Delta^0)^2}$.
Let the special case of $\Omega^0_{0cv}=108.8$ meV and $\Delta^0=-54.4$ meV ---
providing $\Omega^0_N=$ 121.6 meV and $S^0_L/\pi=0.5880 <1$ --- in Fig.~\ref{fig2} be taken into consideration, where the obtained asymmetry parameter is represented as $1/q_a^0\equiv-0.286$.
If the two pulses 0 and 1 generate the same carrier density of $D_L$, the condition of $\delta E_{\boldsymbol{q}} \approx 0$ is realized in both cases, leading to the maximized RZ coupling.
This condition is met by taking, for instance,
$\Omega_N=255.8.0$ meV with $\Omega_{0cv}=244.8$ meV and $\Delta=-74.17$ meV,
which leads to $S_L/\pi=1.3766 >1$.

\begin{figure}[tb]
\begin{center}
\includegraphics[width=7.0cm,clip]{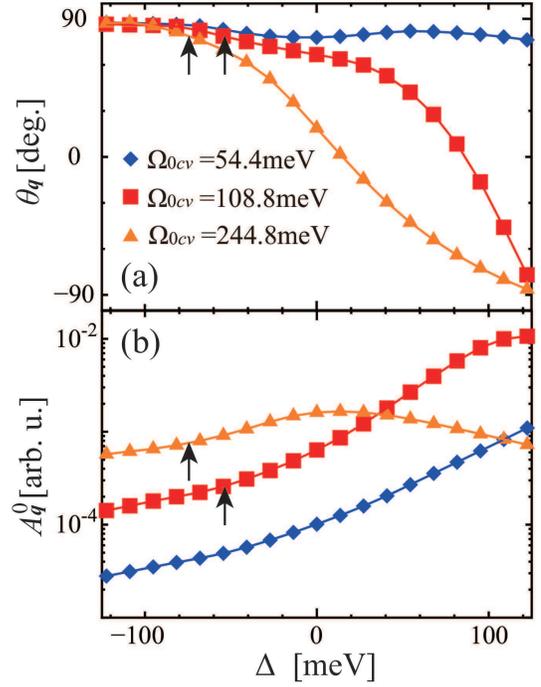}
\caption{The calculated results of (a) $\theta_{\boldsymbol{q}}$ and (b) $A^0_{\boldsymbol{q}}$ as a function of $\Delta$ (in the unit of meV) for $\Omega_{0cv}=$ 54.4 meV (blue diamond), 108.8 meV (red square), and 244.8 meV (orange triangle). As regards the four arrowed values, consult the text for detail.}
\label{fig7}
\end{center}
\end{figure}

Now, the degree of spectral asymmetry $1/q_a$ of the CP generated by pulse 1 is examined.
To this end, we consider the alteration of $\theta_{\boldsymbol{q}}$ and $A^0_{\boldsymbol{q}}$ as a function of $\Delta$, as shown in Fig.~\ref{fig7} for different values of $\Omega_{0cv}$'s of 54.4, 108.8, and 244.8 meV. 
$\theta_{\boldsymbol{q}}$ and $A^0_{\boldsymbol{q}}$ are determined by $\delta V_{\boldsymbol{q}ph}$, as seen in Eqs.~(\ref{theta})-(\ref{deltaV});
$\delta V_{\boldsymbol{q}ph}$ is subject to the photoexcited carrier density, similarly to $\delta E_{\boldsymbol{q}}$.
Thus, with the increase of the carrier density, $(\pi/2-\theta_{\boldsymbol{q}})$ and $A^0_{\boldsymbol{q}}$ tend to increase as a whole in Figs.~\ref{fig7}(a) and \ref{fig7}(b), respectively, aside from the non-monotonic change of $A^0_{\boldsymbol{q}}$ for $\Omega_{0cv}=$ 244.8 meV due to the optical nutation.

The four values arrowed in Fig.~\ref{fig7} represent $\theta_{\boldsymbol{q}}$ and $A^0_{\boldsymbol{q}}$ for $(\Omega_{0cv}, \Delta)=(244.8\:\:{\rm meV}, -74.17\:\:{\rm meV})$ and $(\Omega_{0cv}, \Delta)=(\Omega^0_{0cv}, \Delta^0)$.
Although the RZ coupling is caused in a similar way for both sets of $(\Omega_{0cv}, \Delta)$, the resulting $A^0_{\boldsymbol{q}}$'s are different roughly by three times; $\theta_{\boldsymbol{q}}$'s are almost identical to each other presumably by accident, which equal approximately 80$^{\circ}$.
This difference arises from the fact that $A^0_{\boldsymbol{q}}$ and $\theta_{\boldsymbol{q}}$ are determined through $V_{\boldsymbol{q}ph,ph}^{L\dagger}(0)$ by $\Omega_{cv}(0)$ as well as the photoexcited carrier density, whereas the RZ coupling is determined solely by the latter.
Therefore,  the resulting $1/q_a$-values are not identical in these two $(\Omega_{0cv}, \Delta)$'s of concern, namely, $1/q_a \not= 1/q_a^0$, since the degrees of contribution of $S_{\boldsymbol{q}}^{(out)}(\omega)$ to $S_{\boldsymbol{q}}(\omega)$ are different in these two cases.
However, it is possible that the value of $1/q_a^0$ is retrieved by adjusting the value of $\Omega_N$ so that the condition $S_{\boldsymbol{q}}^{(out)} \ll S_{\boldsymbol{q}}^{(in)}$ is ensured.
Actually, the set of $(\Omega_{0cv}, \Delta)=(313.1\:\:{\rm meV}, -42.78\:\:{\rm meV})$
--- with $\Omega_N=316.0$ meV and $S_L/\pi=1.5$ --- provides the same value of $\delta E_{\boldsymbol{q}}$ as that for $(\Omega_{0cv}, \Delta)=(\Omega^0_{0cv}, \Delta^0)$ and the value of $A^0_{\boldsymbol{q}}$ smaller than that for $(\Omega_{0cv}, \Delta)=(244.8\:\:{\rm meV}, -74.17\:\:{\rm meV})$, though not shown here.
In any case, it would be stated that the spectral asymmetry given by $1/q_a$ varies more or less following the optical nutation with $S_L/\pi >1$.

Finally, two more remarks are given regarding the spectral asymmetry of concern.
First, it is pointed out that the spectral asymmetry attributed to the CP generation is observed exclusively in a lightly $n$-doped Si crystal and semimetals/metals such as Bi and Zn,\cite{hase3,misochko1,hase5,misochko7} though not observed in a GaAs crystal.\cite{ishioka3}
Obviously, the amplitude $A^0_{\boldsymbol{q}}$ of $Q_{\boldsymbol{q}}(\tau)$ in the classical region for GaAs is much greater than that for Si, because in the former, the CP is driven by a much stronger electron-LO-phonon coupling due to the Fr$\ddot{\rm o}$hlich interaction.
Therefore, it is speculated that $S^{(out)}_{\boldsymbol{q}}(\omega)$ dominates $S_{\boldsymbol{q}}(\omega)$ to result in symmetric spectral-profile, that is,
$S^{(out)}_{\boldsymbol{q}}(\omega) \gg S^{(in)}_{\boldsymbol{q}}(\omega)$.
However, it might be possible that $S_{\boldsymbol{q}}(\omega)$ is asymmetric~\cite{Chang2010} when a certain condition leading to $\delta E_{\boldsymbol{q}} \approx 0$ is ensured.
Thus, it would also be necessary to scrutinize spectral profiles at time $\tau$ in the ETR under this condition, for instance, by using the continuous-wavelet transform of $Q_{\boldsymbol{q}}(\tau)$.

Second, it is reported that an asymmetric spectral-profile observed in a lightly $n$-doped Si crystal is attributed to transient Fano resonance.~\cite{hase1,watanabe1,riffe2}
In Ref.~\onlinecite{watanabe1}, based on the PQ model incorporating the Fano model,~\cite{fano} the spectral asymmetry attributed to the Fano resonance is obtained just in the ETR, where $\Omega_{0cv}=16.4$ meV and $\Delta=82$ meV corresponding to $\delta E_{\boldsymbol{q}} \approx -40$ meV; hence, the RZ model is no longer significant.
There, $S_{\boldsymbol{q}}(\omega)$ also shows the asymmetry, since the associated $Q_{\boldsymbol{q}}(\tau)$ tends to vanish rapidly out of the ETR, and thus $S^{(in)}_{\boldsymbol{q}}(\omega)$ becomes dominant over $S^{(out)}_{\boldsymbol{q}}(\omega)$.
Therefore, it is speculated that the asymmetry parameter $1/q_a$ discussed in Figs.~\ref{fig2} and \ref{fig3}(c) becomes finite, namely, non-zero, due to the Fano resonance, even though $|\delta E_{\boldsymbol{q}}|$ increases considerably with $\delta E_{\boldsymbol{q}} < 0$.

\section{CONCLUSIONS}
\label{sec5}

The degree of asymmetry of the spectral profiles of the CP time signals is scrutinized in undoped Si by using the PQ model, and the adiabatic energy curves of the plasmon and the LO-phonon are examined based on the LZ and RZ models.
The asymmetric spectral-profiles are discerned, as long as the RZ interaction is dominant in the ETR with $S^{(in)}_{\boldsymbol{q}}(\omega)$ greater than or comparable with $S^{(out)}_{\boldsymbol{q}}(\omega)$.
Since the magnitude of the RZ interaction depends decisively on the adiabatic plasmon energy, thus the degree of asymmetry varies with respect to the photoexcited carrier density.
Actually, it is confirmed in both of the present calculation and experiment that the spectra become more symmetric with larger 
change of the carrier density from the critical carrier density at which the effect of the RZ interaction is maximized.

The particular emphasis is put on the degree of this asymmetry in both of the far-lower carrier density region ($\omega_{\boldsymbol{q}pl} \ll \omega_{\boldsymbol{q}}$) and far-higher carrier density region ($\omega_{\boldsymbol{q}pl} \gg \omega_{\boldsymbol{q}}$) than the region concerned in this study ($\omega_{\boldsymbol{q}pl} \approx \omega_{\boldsymbol{q}}$).
In the lower carrier density region, it is reported that the transient Fano resonance with an asymmetric profile is discerned in the ETR;~\cite{watanabe1} 
this is caused by the interaction between the LO-phonon mode and the single-particle excitation mode, and there is little contribution from the plasmon mode.
On the other hand, in the higher carrier density region where pulse areas of the corresponding irradiated laser pulses exceed $\pi$, it would be possible that the degree of asymmetry of the profile varies following the optical nutation of the photoexcited carrier density.
Here, the RZ interaction between the LO-phonon mode and the plasmon mode in the ETR still plays a significant role of determining the degree of asymmetry.
These two phenomena occurring in the regions of $\omega_{\boldsymbol{q}pl} \ll \omega_{\boldsymbol{q}}$ and $\omega_{\boldsymbol{q}pl} \gg \omega_{\boldsymbol{q}}$ are still unexplored in experiments.
Therefore, it is expected to measure the spectral profiles of CP over the wider region of photoexcited carrier density than the present, and to reveal the underlying quantum-mechanical effects governing the alteration of the profiles.

\begin{acknowledgments}
This work was supported by JSPS KAKENHI Grants No. JP23540360 and No. JP15K05121, NSF Grants DMR-0116034 and OISE-1743717, and NIMS Research Funds.
\end{acknowledgments}


\end{document}